\documentstyle[11pt,aaspp4]{article}  % For preprint formating
\newcommand{\kms}{\mbox{ km~s$^{-1}$}}
\newcommand{\kmm}{\mbox{ km~s$^{-1}$ Mpc$^{-1}$}}
\newcommand{\etal}{\mbox{ et~al.}}

\newcommand{\mgib}{\mbox{ Mg$\,$I$\,$b}}
\newcommand{\cri}{\mbox{ Cr$\,$I 5208}}
\newcommand{\fei}{\mbox{ Fe$\,$I 5227}}
\newcommand{\cai}{\mbox{ Ca$\,$I$+$Fe$\,$I 5270}}
\newcommand{\ql}{\mbox{$0957+561$}}

\newcommand{\qp}{\mbox{Q$\,0957+561$}}
\newcommand{\Ho}{\mbox{H$_\circ$}}
\newcommand{\tba}{\mbox{$\Delta\tau_{\rm BA}$}}

\begin{document}
%{\centerline {\bf DRAFT -- NOT FOR DISTRIBUTION}}
\title{An estimate of \Ho\ from Keck
spectroscopy of the gravitational lens system \ql}

\author{Emilio E. Falco \& Irwin I. Shapiro}
\affil{Harvard-Smithsonian Center for Astrophysics\\
60 Garden Street, Cambridge MA 02138}
\and
\author{Leonidas A. Moustakas \& Marc Davis}
\affil{Astronomy Department, University of California, Berkeley, CA 94720}

\begin{abstract}
We present long-slit LRIS/Keck spectroscopic observations of the
gravitational lens system \ql.  Averaged over all of our data, the
rest-frame velocity dispersion $\sigma_v$ of the central lens galaxy
G1 is $\sigma_v = 279 \pm 12$ \kms. 
However, there appears to be a
significant decrease in $\sigma_v$ as a function of distance from the
center of G1 that is not typical of brightest cluster galaxies. 
Within 0\farcs2 of the center of G1, we find 
the average $\sigma_v = 316 \pm 14$ \kms, whereas for positions 
$>0\farcs2$ from the center of
G1, we find the average $\sigma_v= 266 \pm 12$ \kms. A plausible
explanation is that G1 contains a central massive dark object of mass
$M_{MDO} \approx 4\times 10^{9} h_{100}^{-1} {\rm M}_\odot$ ($h_{100} =
\Ho/100\kmm$), which contributes to the central velocity dispersion,
and that the outer value of $\sigma_v$ is the appropriate measure of
the depth of the potential well of G1.  The determination of a
luminosity-weighted estimate of $\sigma_v$ is essential for a
determination of \Ho\ from \qp; our accurate measurements remove one
of the chief uncertainties in the 
assumed form of the mass distribution of the lens. 
Thus, with the
recent apparent reduction in the uncertainty in the
measurement of the time delay for the images A and B of \qp, 
$\tba = 417 \pm 3$ d (Kundi\'c et al. 1996), we obtain
an estimate for the Hubble constant: $\Ho = 62\pm7$ \kmm.  
If for some reason the trend of $\sigma_v$
with slit position is spurious and we should use the dispersion
averaged along the slit, then the estimate of \Ho\ increases 
to $67\pm8$ \kmm.
These standard errors do not, however, include any contribution
from any errors in the assumed form of the mass distribution of the
lens. In particular, we used the mass model described by 
Falco, Gorenstein \& Shapiro (1991), as updated by Grogin 
\& Narayan (1996a, b). 
The reduced $\chi^2$ of model fits to the available position
and magnification data for this system is relatively high ($\sim
4$), indicating that 
the estimate of \Ho\ may have a significant contribution from model
errors. Further observations, discussed herein, should allow such errors to
be estimated reliably. 

\end{abstract}

\section{Introduction}

Multiply-imaged quasars are excellent laboratories for the estimation
of cosmological parameters, particularly \Ho.  Lens systems showing
time variability are essential, since measurements of time delays for
lensed image pairs are needed to scale from redshifts and angles to
physical length scales, as first discussed in two seminal papers by
Refsdal (1964, 1966).

The gravitational lens system \ql\ was the first to be discovered
(Walsh, Carswell, \& Weymann 1979).  It is observed to consist of two
images A and B of a QSO at $z=1.41$, and a lens containing primarily a
cluster of galaxies at $z = 0.36$ (Young \etal\ 1981;
Angonin-Willaime, Soucail \& Vanderriest 1994, hereafter ASV94;
Garrett, Walsh \& Carswell 1992). This cluster is thought to be
responsible for the wide ($\sim 6\arcsec$) separation of the images.
The brightest member of the cluster is a giant elliptical galaxy, G1,
only $\approx 1\arcsec$ North of image B (Stockton 1980). Because it
is bright, and therefore presumably massive, and near B, G1 plays a
critical role in the lensing of \qp. There is also evidence that
another cluster of galaxies ($z \sim 0.5$) 
lies along the line of sight to the QSO (ASV94); 
at the present level of accuracy of the data and of
the lens models, ignoring the fact that this cluster is at a 
different redshift has a negligible effect on our estimate of \Ho\ 
(also, see Fischer \etal\ 1996).

Both QSO images have strong radio emission; 
VLBI brightness maps exhibit single jets
whose positions, structures and orientation angles have been accurately
determined (Gorenstein \etal\ 1988b; Garrett \etal\
1994; Campbell \etal\ 1994). The QSO images also radiate in X-rays 
and exhibit there rather dramatic 
time variability (Chartas \etal\ 1995). The
images of \qp\ have been closely monitored for radio (Leh\'ar \etal\
1992; Haarsma \etal\ 1996) 
and optical (Schild \& Cholfin 1986; Vanderriest \etal\ 1989;
Schild 1990; Schild \& Thomson 1995; Kundi\'c \etal\ 1995, 1996)
variability since their discovery.  Their light curves reveal
variability with modest amplitudes ($< 20\,$\% over the previous
decade), and there has been a continuing controversy regarding the
correct estimate from these variations of the time delay \tba\ for A
and B.  Such estimates have been either approximately 1.5 yr (Press,
Rybicki \& Hewitt 1992a, 1992b), or 1.1 yr (Schild 1990; Schild \&
Thomson 1995; Pelt \etal\ 1996), with the most recent 
evidence now favoring the latter value
(Kundi\'c et al. 1996; Haarsma \etal\ 1996).

Detailed modeling of \ql\ has been extensively carried out in recent
years (Falco, Gorenstein \& Shapiro 1991, hereafter FGS91; Grogin \&
Narayan 1996a, hereafter GN96a, see also an erratum: Grogin \& Narayan
1996b, hereafter GN96b).  The modeling results point clearly to two
types of measurements required to convert an observed time delay into
a value of \Ho: the central velocity dispersion $\sigma_v$ of G1 and
the surface mass density of the cluster (FGS91; GN96a). These
measurements are complementary and can be used to break a degeneracy in
lens models that describe the cluster and G1 (Gorenstein, Falco \&
Shapiro 1988a).  Eliminating this degeneracy can lead to a robust
estimate of \Ho. However, systematic errors may affect the 
conversion of $\sigma_v$ to a lensing mass, as 
discussed by Kochanek (1991) and in GN96a. For example, anisotropy 
in the velocity distribution of stars in G1 affects the relationship
between observed dispersion and actual dispersion. 
Measurements of the velocity dispersion and of
the light distribution of G1 with much higher SNR 
than presently available would be required to study these
effects.

We have concentrated on the measurement of $\sigma_v$, for which Rhee
(1991) obtained a value of $303 \pm 50$ \kms, with the 4.2$-$m William
Herschel Telescope.  The measurement is difficult due to the proximity
of image B to G1, and to the relatively high redshift of G1.  We took
advantage of the large collecting area of the 10$-$m Keck I telescope
and of the capabilities of the Low-Resolution Imaging Spectrograph
(LRIS; Oke \etal\ 1995), a combination that proved sufficient to
obtain a useful new result, as we report below. We used the same
absorption line triplet, \mgib, as Rhee (1991), because it is
relatively free of contamination from QSO and telluric lines
(Stevenson 1994).

We describe our observations and our calibration of velocity
dispersion measurements in \S 2. In \S 3, we conclude with a
discussion of our results and their significance regarding estimates
of \Ho.

\section{Observations}

We obtained spectra of G1, three template stars (bright K giants), and
M81 with LRIS on Keck I on 1995 March 28.  We used the 600 g~mm$^{-1}$
grating (blazed at 7500 \AA) with a $1\arcsec\times180\arcsec$ slit
(the same slit width as that of Rhee 1991) for all our
observations. The spectral coverage was 5098$-$7696 \AA\ for the
stellar templates and M81, and 6149$-$8747 \AA\ for G1. The latter
range includes the \mgib\ absorption triplet with rest $\lambda
\approx 5175$ \AA, which corresponds to $\lambda \approx 7038$ \AA\
for $z \approx 0.36$.  The detector was a Tektronix CCD with 2048$^2$
24$-\mu$m pixels, each subtending $0\farcs213$ on the sky (Oke \etal\
1995); the corresponding dispersion was 1.3 \AA/pixel.  Table 1 shows
a log of the observations, listed in the order of their
acquisition. The table includes the seeing for each of our exposures,
which we estimated as the mean of three measurements of the spatial
FWHM of unresolved objects (the A image and a nearby star), one each
at the red and blue ends of the spectrum, and one at intermediate
wavelengths.  Figure 1 is a diagram of the orientation of the slit
during our exposures.  In addition to the object exposures, we
obtained calibration arc spectra with Hg-Ne-Ar lamps, as well as
internal flats and biases, as usual for CCD calibrations. The slit and
spectrograph combination yielded a spectral resolution of $5.9 \pm
0.2$~\AA, estimated as the mean FWHM of isolated lines in our
calibration spectra.

We tested for measurable effects of flexure in exposures 1$-$4 by
measuring the position of the peaks and the FWHM of unresolved sky
emission lines at 6365~\AA, 6950~\AA, 7370~\AA, and 7750~\AA.  We
determined centers and FWHM for these lines at three different spatial
positions along the slit, one at the midpoint of the A$-$B line (which
is nearly N$-$S), and one each $\sim 10\arcsec$ N and S of this
location. We found that the line centers shifted from exposure to
exposure by $< 1$~\AA, and by $\sim 2$~\AA\ from exposure 1 to
exposure 4. The mean FWHM of the lines within each exposure was again
$\sim$ 5.9~\AA, identical to the width of the comparison lines
obtained with very short exposures.  If the lines were broadened by
1~\AA\ flexure during each exposure, the resulting FWHM of sky lines
would be changed by too small a value for us to detect.  A 2~\AA\
broadening would have resulted in sky linewidths of order 6.2~\AA,
marginally inconsistent with the measured linewidths.  If our data
were contaminated by 1~\AA\ flexure, a 32 \kms\ width should be
subtracted in quadrature from the measured rest frame velocity
dispersion; for the measured dispersion derived below such
contamination would make only a 2 \kms\ difference in our final
answer.

The CCD frames were processed with the LRIS software package developed
by Drew Phillips (Lick Observatory), for bias subtraction and
flat-fielding (only the long G1 exposures were flat-fielded), taking
into account the two-amplifier readout mode that we used with the
CCD. We reduced the spectra with the standard IRAF\footnote{IRAF
(Image Reduction and Analysis Facility) is distributed by the National
Optical Astronomy Observatories, which are operated by the Association
of Universities for Research in Astronomy, Inc., under contract with
the National Science Foundation.}  ``longslit'' package. We derived a
wavelength solution from the arc exposures (using task ``identify''
along CCD rows, and fitting Legendre polynomials of up to 4th order),
and determined the spatial distortion of the spectra from the object
exposures (using tasks ``identify'' and ``reidentify'' along CCD
columns). We calculated wavelength calibration and spatial distortion
solutions with task ``fitcoord'' (using Chebyshev polynomials of 4th
and 3rd order along CCD lines and columns, respectively), and
rectified the spectra with task ``transform.'' Part of the processing
with ``transform'' included rebinning the spectra both linearly and
logarithmically along the spectral direction (the latter for ultimate
analysis using the Fourier cross-correlation method, as we describe
below).

For the G1 exposures, the slit was always centered on the B image of
the QSO, in an attempt to obtain simultaneously the best possible
measurement of the spectrum of the QSO. In exposures 1$-$3, the slit
was oriented N$-$S (PA = 180$^\circ$ for exposure 3), while in
exposure 4, the slit was tilted $11^{\circ}$ W of N$-$S, to obtain the
best possible spectrum of image A.  However, exposures 1$-$3 still
include a significant amount of light from A. We found the separation
for QSO images A and B on exposure 4 to agree well with the previous
VLBI measurement (Gorenstein \etal\ 1988b) at the $\sim 0\farcs01$ level,
thus confirming our choice of slit orientation at the $\sim 3^\circ$ level.

We subtracted the sky background from exposures 1$-$4 with task
``background.'' The spatial distortion of the spectra was significant;
therefore we fit the sky only in two 15-pixel fitting strips, one just
below B and one just above A, and fit a slope to the background in the
spatial direction of the two strips. Because we are interested in the
\mgib\ triplet, the complicated structure of OH emission in the
neighborhood of 7000 \AA\ made it essential to subtract the sky
accurately in the $\pm 10\arcsec$ vicinity of B. We determined the
quality of our sky fits by checking the residuals as a function of
distance from B to its South, and from A to its North. Figures 2 and 3
show sky spectra extracted from exposure 1 at $\sim 9\arcsec$ and
$\sim 4\arcsec$ S of B, as well as the residuals after the same sky
subtraction procedure that we applied to our object spectra; 
these are typical of our results. 
The magnitude of the residuals is consistent with our noise 
level, except where strong sky lines are present; such regions are
excluded from our velocity dispersion estimates.
We found that, as can
be seen in the figures, the sky spectra are dominated by OH line
emission.  There is no significant telluric absorption in the region
of interest.  Outside of the OH bands, the sky-subtracted spectra have
noise per pixel that is approximately equally divided between sky
noise and detector noise.  (The amplifier gain is approximately
1.6~e$^-$~DN$^{-1}$ and the readout noise is approximately 8~e$^-$.)
In the spatial region of chief interest, the sky subtraction is
acceptable even throughout the OH emission band. We tried subtracting
a parabola instead of a slope, and narrowing or widening the fitting
strips, but found no significant change in the quality of the
subtraction. For estimation of the velocity dispersion described
below, we worked in the narrowly defined band, $6912-7220$ \AA, to
avoid the complications of the poorly subtracted OH bands outside this
range.
 
Figure 4 shows spectra of B and A extracted from exposure 4, which was
aligned with the B$-$A line, and therefore yielded the QSO spectra of
highest signal to noise ratio (SNR) in our set.  Table 2 shows a list
of narrow absorption lines in our spectra of B and A.  These lines, as
well as a broad Mg II emission feature, had been detected previously
(Weymann et al. 1979; Wills \& Wills 1980).  Our measurements of
absorption line centers and equivalent widths (with the spectral-line
measurement facilities in ``splot'' within IRAF) agree within 
the standard error ($\sim 0.6$ \AA\ RMS, or $\sim 26$ \kms\ at 7000~\AA)
with the previously
published estimates.  Our measurements of emission line centers are
more uncertain ($\sim 5$ \AA\ RMS) because the Mg II absorption lines
in the wings of the emission add significant uncertainty.  The spectra
also show possible blends of Fe II emission from the QSO, at
$\lambda_{obs} \sim 6300$ and $7140$ \AA.

Figures 5$-$8 show spectra that we extracted from exposures 1$-$4 from
single rows in a strip of the CCD that, 
projected on the sky, corresponds to separations
between each row and the optical center of G1 
(Stockton 1980) from $\sim -$0\farcs4 to $\sim +$0\farcs5, negative (positive)
signs indicating South (North) of this center.  
We selected CCD rows starting from
the row in which visual inspection just revealed the presence of a
\mgib\ line, and ending with the row just before the one for which
visual inspection failed to detect the same line. 
We registered 
the different exposures with one another by assuming that the
CCD row with the largest peak brightness contained
the center of brightness of image B. A continuum was
subtracted from each of these spectra by fitting cubic splines of
order 16, with 5 iterations to reject points with residuals higher
than $2 \sigma$ and lower than $5 \sigma$ from the fit, where $\sigma$
is the RMS dispersion of each fit.  The figures show the spectra after
the continuum subtraction.  
For each exposure, the total number of CCD
rows that yielded a useful spectrum (in the sense defined below) was
$5$, for a total of $20$ useful spectra that we will refer to
hereafter as ``G1 spectra.''  Telluric O$_2$ absorption is apparent in
bands at 6870~\AA\ and at 7600~\AA, while telluric H$_2$O absorption
is present at 7240~\AA\ (Stevenson 1994).  Fortunately, the spectral
features of interest for G1 do not fall on any of these bands and
therefore no correction was made for these absorption features. We
checked that no telluric absorption is present in our spectra in the
region of interest, by examining spectra extracted from the slit on
the North side of image A; we are fortunate that this region is clean.
In the vicinity of \mgib, the G1 spectra reveal additional absorption
lines, which can be identified with \cri ~\AA, \fei ~\AA\ and \cai
~\AA. These absorption lines affect our measurements at the $\sim 7$
\kms\ level (see our discussion of calibration of the velocity
dispersion, below).

Note from Figures 5$-$8 that the G1 spectra extracted in the first row
($\sim 0\farcs4$ from the optical center of G1) 
are strongly contaminated by QSO emission
that has leaked through the spline continuum fit.  We experimented
with subtracting a scaled version of the spectrum of image A from the
individual G1 spectra, and found that such a procedure cleanly
eliminates the contribution of image B to the region centered at
6750~\AA, but that, as expected, the resulting G1 spectra have
measurably increased noise.  Because the spectrum of the QSO is
featureless in the narrow region of interest, we decided against doing
this subtraction in the final analysis.

We calculated the redshift of G1 and its line-of-sight velocity
dispersion using the Fourier cross-correlation method (Tonry \& Davis
1979), as implemented in the IRAF task ``fxcor.''  Given spectra of an
object and of a template, fxcor yields estimates of the redshift of
the object and of the FWHM of absorption lines in the object
spectrum. These quantities are estimated by locating the peak in the
cross-correlation and fitting a gaussian function to it, after
background subtraction. The redshift of the object spectrum is then
inferred from the location of the peak of the gaussian fit to the
cross-correlation; the corresponding estimate of velocity
dispersion is related to the FWHM of the gaussian, as we describe
below.

Our strategy with fxcor was to: (1) subtract continua from the spectra
of the templates, in the same fashion as for the G1 spectra; (2)
restrict the region of the spectrum of G1 included in the
cross-correlation to $\sim 6900 - 7220$~\AA\ and that of the templates
to $\sim 5090 - 5330$~\AA, to avoid contributions from the QSO and
poorly-subtracted telluric emission lines; (3) apodize the spectra by
5\% (at each end) to minimize spectral aliasing effects; (4) restrict
the gaussian fits to the central 21 pixels of the cross-correlation
peak, to minimize the influence of the noise background (varying this
fitting width 
between 9 and 27 pixels affects our estimates by $\leq 5$ \kms;
we include this value in our error budget
as a systematic contribution); (5) filter out the lowest
frequencies (wavenumbers $\leq 130$ \AA$^{-1}$; compare with Franx,
Illingworth \& Heckman 1989) present in the Fourier-transformed
template spectra, to remove any remaining low-frequency background.

Our template spectra and galaxy spectra have different resolutions as
measured in \kms\ (see, e.g., van Dokkum \& Franx 1996).  The mismatch
occurs because the resolution of the spectrograph is nearly constant
in \AA\ and the template is at rest, whereas G1 is redshifted to
$z\approx 0.36$.  Thus, the resolution in \kms\ of our template
spectra (FWHM $\sim 340$ \kms\ at 5200 \AA) is coarser than that of
our G1 spectra (FWHM $\sim 250$ \kms\ at $\sim 7072$ \AA) by a factor
of $\sim 1.36$.  To allow us to estimate the effect of such a mismatch
on our estimates of velocity dispersions, M. Franx kindly provided
spectra of template stars that D. Fisher (U.C. Santa Cruz) 
acquired with the Kast
Spectrograph on the Lick Observatory 3$-$m telescope on 1993 May 16.
These spectra covered the range from 4214 to 5620 \AA, with resolution
$\sim 186$ \kms\ at 5200 \AA, $\sim 1.3\times$ the resolution of our
G1 spectra. Thus, we use these spectra to avoid the resolution
mismatch.  We selected spectra of three of the stars, HD126778,
HD172401 and HR4435, where the \mgib\ line was as prominent as in our
LRIS templates.  We followed the procedure of van Dokkum \& Franx
(1996), wherein before cross-correlation, the template spectra are
rebinned and smoothed to match the resolution of the G1 spectra in
\kms.

We obtained LRIS and Lick average templates by summing the
corresponding spectra (both smoothed and unsmoothed in the Lick case).
We calibrated our calculations of velocity dispersions by convolving
the LRIS and Lick average template spectra with a series of 11
gaussians with $\sigma_v$ increasing in steps of 50\kms, from 0 to
500\kms.  We then ran fxcor for each of these, with the same
parameters and wavelength sampling as those for G1. Figure 9 shows a
calibration plot of the FWHM of the output cross-correlation peak as a
function of ``input'' velocity dispersion, both in \kms,
for the LRIS and Lick templates.  The velocity dispersion $\sigma_v$
of G1 can be obtained from such curves, by first blueshifting the
galaxy spectra to $z=0$ (see below for our estimate of $z_{G1}$), then
running fxcor on these rest spectra, and finally by interpolating the
calibration curves.  The plot shows calibrations for the averages of
our LRIS templates and of our Lick templates, both smoothed and
unsmoothed. The figure shows that our procedure is indeed sensitive to
the mismatch in resolutions of the spectra of the templates and of
G1. As can be seen in Figure 9, for the average G1 spectrum and the
average LRIS template (unsmoothed average Lick template), the
``output'' FWHM of the cross-correlation peak is $\sim 813$ \kms
($\sim 662$ \kms), and the corresponding ``input'' $\sigma_v$ is $\sim
238$ \kms ($\sim 287$ \kms). For the smoothed average Lick template,
the ``output'' FWHM and input $\sigma_v$ are $\sim 683$ \kms and $\sim
279$ \kms, respectively.  We adopt the curve for the smoothed average
Lick template as the calibration of our velocity dispersions.

We summed the individual rows with G1 spectra to maximize the SNR.
Thus, we obtained a mean G1 spectrum, as well as averages by exposure
and by row, to estimate the RMS spread of our measurements.  We also
repeated the velocity dispersion estimation procedure using in turn
each of the 3 stars in Table 1 as a template, with the G1 spectra in
their role of object spectra.  For each template, we found the mean
and the RMS spread of the FWHM estimate for the mean G1 spectrum.  For
each of the four separate exposures, we added the data from the five
useable rows of CCD data.  Table 3 lists the derived $\sigma_{v}$ of
the individual exposures relative to the mean of the three template
stars.  We list the measured FWHM of the cross-correlation routine,
and the inferred $\sigma_{v}$ for each measurement, as well as the
$R-$value, indicative of the SNR for each cross-correlation peak that
we used (Tonry \& Davis 1979).  The largest source of uncertainty in
our velocity estimates is the variation from exposure to exposure.
The mean $\sigma_{v}$ averaged over the four exposures is $279 \pm 9$
\kms, where the error is the standard error of the mean of the four
independent observations.  This error represents a fair estimate of
the statistical uncertainty of our velocity dispersion measurement.

Also listed in Table 3 are the 
results from the cross-correlation of the sum
of all 20 rows of G1 data (shown in Figure 10; see also Figure 11)
and each of the 3 template stars.  Again, the results are
very consistent, and the standard deviation of the mean of $\sigma_v$ 
for G1 is 2 \kms.  We tested for the effects of the \cri ~\AA,
\fei ~\AA\ and \cai ~\AA\ lines in the wavelength regions that we used
for cross-correlations. We found that restricting these regions even
further, to include only the \mgib\ triplet, changed our mean velocity
dispersion estimate to $\sim 272$ \kms. Thus, the presence of these
lines adds (in quadrature) a $\sim 7$ \kms\ standard error to our error
budget. These are reasonable estimates of the {\it systematic} errors
due to the possible mismatch of the template spectrum to the
intrinsic spectrum of G1.  We sum all the entries in our error budget
in quadrature (see Table 4), arriving at a final estimate of
$\sigma_{v} = 279 \pm 12$ \kms. However, this standard error does not 
include any contribution from a possibly spatially-dependent variation in
the mass-to-light ratio in G1. 

Finally, Table 3 lists the results from the 
cross-correlation with the average of each of the
five rows with useable spectra that we extracted from our exposures.
We summed each row over all four exposures to calculate the
cross-correlations, to check whether our data yielded a detected
variation of the velocity dispersion of G1 as a function of the
distance of the strip from the center of brightness of G1, which we
assumed to be at the Stockton (1980) coordinates as measured from
image B.  Table 3 and Figure 12 show a 
seemingly very significant trend with
position from the center of G1.  The $\chi^2$ fit for a null
hypothesis, that all the dispersion measures are consistent with a
constant value, is $\sim 22.2$ for 4 degrees of freedom and 
absent significant systematic error, can be
ruled out with high confidence.  For rows 2 and 3, both within
0\farcs2 from the center of G1, we measure a mean $\sigma_{v} = 316
\pm 14$ \kms, whereas for rows 1, 4, and 5, all at more than 0\farcs2
from the center of G1, we measure $\sigma_{v} = 266 \pm 12$ \kms\
(including all sources of error).  The average $\sigma_{v}$ derived
from the 5 rows is $\sim 286$ \kms, larger than $\sigma_{v}$ averaged
over exposures, but within our standard errors.

With fxcor, we also measured the redshift of G1: $z_{G1} = 0.356 \pm
0.002$.  Our average template spectrum is compared with that of the
average of G1 in Figure 10, after convolution with a gaussian with
$\sigma = 279$ \kms.  As an additional reference, in Figure 11, we
plot the spectrum of the smoothed average Lick template, both before
and after convolution with the gaussian.

As a partial check of our procedure, we determined the central
velocity dispersion of M81, $\sigma_v = 180 \pm 10$ \kms, which
compares well with previous estimates (e.g., $177 \pm 13$ \kms; Keel 1989), 
but which
does not suffer from the resolution mismatch that we had for G1. We
also utilized the program ``four'' kindly provided by M. Franx to
obtain independent estimates of $\sigma_v$. 
These estimates agreed with ours to one more figure than is significant.

\section{Discussion and Conclusions}

As first shown in Falco, Gorenstein \& Shapiro (1985; see also FGS91
and GN96 a,b), a degeneracy exists in lens models, between mass in the
cluster and in G1. Any lens model can be transformed by adding a sheet
with uniform convergence $\kappa < 1$ to the cluster, and reducing the
mass of G1 by a factor $(1 - \kappa)$. Such a transformation maintains
the observables invariant, except for \tba.  Thus, by obtaining a
value for the mass of G1, one is able to break the degeneracy, and
with an estimate for \tba, to obtain an estimate of \Ho.  Because the
mass of G1 can be estimated from $\sigma_v^2$, our measurement breaks
the degeneracy. 
However, the models of FGS91 and GN96a, b assume that there is no significant 
additional variation in shear on the 6\arcsec\ scale of the separation of 
the images; as noted below, this assumption may be somewhat in error. 

As described by FGS91 and GN96a, and as updated by GN96b, the 
model-fitting now favors the FGS91 mass model, which yields 
the following estimate of \Ho\ for the \qp\ system: 
$$\Ho = 98^{+12}_{-11} \left({\sigma_{v}\over 330 ~{\rm \kms}}\right)^2
\left({1.1 ~{\rm yr}\over \tba } \right) {\kmm},$$ where the errors
are $2\,\sigma$ bounds, as defined in GN96b, and do not
include any contribution from the uncertainty
in the mass model.  In the case
$\sigma_{v} = 330$ \kms, the galaxy would provide all the convergence,
while the cluster would provide none.  
Our average rest-frame velocity dispersion
estimate, $\sigma_{v} = 279 \pm 12$ \kms, is consistent with the value
obtained by Rhee (1991) with a slit of the same width as ours, but
with significantly lower SNR.
The most recent estimate of \tba\ is $417 \pm 3$ d (Kundi\'c \etal\ 1996). 
Using our average value of
$\sigma_{v}$ and that of Kundi\'c et al. (1996) for \tba\ 
yields an estimate for the Hubble constant of $\Ho =
67\pm8$ \kmm, where the error now
includes the total observational uncertainty, but {\it not} the 
model uncertainty.
This result appears to be competitive 
in accuracy with those obtained from observations of 
objects at lower redshifts. 
 
The magnitude of the measured average decrease of $\sigma_{v}$ with angular
distance from the center of G1 (both toward and away from image B)
possibly suggests the existence of a MDO within G1, but data of 
significantly higher angular resolution is needed 
(recall that $z_{G1} = 0.36$) 
before one can estimate parameters for the  MDO.
The trend of $\sigma_v$ does not seem quite consistent with the gradual slope
observed in brightest cluster galaxies 
($\Delta \log\sigma_v/\Delta \log r = -0.061 \pm 0.004$; 
Fisher \etal\ 1995),  and neither is the trend consistent
with the expected signature of an MDO.  If the luminosity distribution of G1
were sufficiently ``cuspy'', our spectra, taken in conditions of $\sim
0\farcs7$ seeing, might be contaminated by the increased dispersion
this central mass would introduce.  
If we take as a (crude) measure of the mass of the central MDO the
virial mass corresponding to (a) the velocity dispersion, $\sigma_{MDO}$,
obtained by the quadrature subtraction of the average 
outer dispersion from the average inner dispersion; 
and (b) the assumption that
the size of the relevant light distribution contributing to the
inner dispersion is 0\farcs2 ($\approx 0.6h_{100}^{-1}$~kpc at G1), 
we obtain $M_{MDO} \approx 4\times 10^{9}h_{100}^{-1} {\rm M}_\odot$, because
$\sigma_{MDO} = 171\pm 20$ \kms.  
Note that the FGS91 model (as updated in GN96b)  
also suggests the existence of an MDO within G1, but 
the inferred $M_{MDO}$ is negligible compared to the mass required to
explain the \qp\ image splitting.
Note also that our $M_{MDO}$ estimate is comparable to those 
inferred for MDOs within nearby galaxies such as M87 
($\sim 2.4\times 10^{9}{\rm M}_\odot$; Ford
\etal\ 1994) and NGC3115 ($\sim 10^{9}{\rm M}_\odot$; 
Kormendy \etal\ 1996). The estimate for $M_{MDO}$ may 
be considered an upper limit, because the light of G1 may well be more
concentrated than we assumed; the estimate might be improved with an
HST image of the galaxy (which recently became available from 
the HST archives). Thus, it may not be appropriate for
estimation of \Ho\ to use the average $\sigma_{v}$, if it is
influenced appreciably by an MDO in the nucleus.  Assuming that rows
1, 4, and 5 yield the appropriate measure of the velocity dispersion, 
we find that $\sigma_{v}= 266 \pm 12$ \kms\ corresponds to the 
total mass of G1, and that $\Ho = 62\pm7$ \kmm.  

An accurate measurement of \Ho\ is fundamental for progress in
cosmology, and the subject is alive with activity.  Our estimate of
\Ho\ is consistent with results emerging from a
number of separate estimates based on widely differing methods (see,
e.g., proceedings of the May 1996 conference on the ``Extragalactic
Distance Scale" at STScI); it also agrees with the rather broad limits 
set by recent measurements of time delays for the PG
1115$+$080 quadruple gravitational lens system 
(Schechter et al. 1996); 
more recently Keeton \& Kochanek (1996) have inferred from this
system a value of $\Ho = 60 \pm 17$\kmm.

The cluster's convergence and the line-of-sight velocity dispersion of
G1 are measured independently, and so should give self-consistent
values, in the sense that $1-\kappa = (\sigma_v/330~{\rm \kms})^2$.
The cluster convergence we infer 
for $\sigma_{v}= 266 \pm 12$ \kms\ is $\kappa=0.35\pm0.06$; 
for $\sigma_{v}= 279 \pm 12$ \kms, we find $\kappa=0.28\pm0.06$.
These values are consistent with the value found by Kundi\'c et
al. (1996), using the cluster mass distribution modeled by Fischer 
\etal\ (1996), $\kappa = 0.24 \pm 0.14 (2\sigma)$.
 
Mapping the shear field of the cluster with HST by measuring the
distortions it induces on the background field of galaxies, could lead
to a better estimate of its surface mass density (see, e.g., Tyson, Valdes
\& Wenk 1990; Tyson \& Fischer 1995; Schneider \& Seitz 1995; Seitz \&
Schneider 1995; Squires \etal\ 1996a, b; Seitz \& Schneider 1996), and
thus provide an additional check of the cluster convergence we infer.
Such a test is currently underway (Rhee \etal\ 1996).

To help remove remaining model uncertainties, other 
measurements of the \qp\ system would 
also be useful.  For example, our observation
should be repeated on a night with better seeing, and 
in addition observations should be made with an E$-$W
orientation of the slit.  A high-quality image in the
optical or near-IR would be useful to judge the ``cuspiness'' of G1
and to assess whether it has a central MDO or even a black hole.

A more reliable measurement of the velocity dispersion of the cluster
would also be useful to refine estimates 
of its virial mass; we have acquired spectra of $\sim 25$ additional
galaxies in the field of \qp, which we plan to combine with the
redshift measurements of ASV94 to improve the estimate of the cluster
velocity dispersion.

Improved modeling must also be pursued for the \qp\ system.  As
noted above and emphasized by GN96a, b, the poor reduced $\chi^2$ of the 
models is a cause for concern
(see also Kochanek 1991).  The VLBI plus optical data 
now yield, in effect, 11 
observational constraints; the models already include 5 
parameters (3 for the G1 galaxy, plus 2 for the cluster),
leaving only 6 degrees of freedom.  Introduction of additional
parameters via a more complex model will further reduce the number of
degrees of freedom, so one must be very judicious  in
the interpretation of these more complex models,
because a large number of parameters could be used
to fit any constraint, but their values would not necessarily indicate a
meaningful range of physical properties. 

Until we have obtained more data
(and completed the reduction of others), as described above, and examined our
models of \qp\ in greater depth, 
this tantalizing measurement of the fundamental
cosmological parameter \Ho\ must be considered 
only suggestive and not definitive.

\bigskip
\noindent {\bf Acknowledgments} 

The W. M. Keck Observatory is a scientific partnership between the
University of California and the California Institute of Technology,
made possible by the generous gift of the W. M. Keck Foundation.  This
work was supported in part by NSF grants AST92-21540 and AST93-03527.
We thank M. Franx, C. Kochanek, M. Geller and S. Kenyon for useful
comments.  LM would like to thank SAO for travel support to the CfA.
We also thank the referee, Tomislav Kundi\'c, for insightful comments that
helped us clarify our results.

\newpage

\section{FIGURE CAPTIONS}

\noindent Fig. 1: The \qp\ field, with labels indicating the images A
and B, and the galaxy G1. The diameters of the circles are $\sim
0\farcs9$, corresponding to our worst seeing (Table 1).  Vertical
solid lines represent the slit for exposures 1$-$3; dotted lines at an
angle of $\sim 11^{\circ}$ from vertical represent the slit for
exposure 4. Horizontal dashed lines show the positions of the CCD rows
that we used in our analysis, as they would appear if they were
projected on the sky.

\noindent Fig. 2: Sky spectrum $9\arcsec$ S of B (heavy 
curve) and residuals (light curve) 
after the usual sky subtraction (see text).  The label indicates the
location of the \mgib\ triplet, redshifted to $z \sim 0.36$.

\noindent Fig. 3: Same as Figure 2, except for $4\arcsec$ S of B. 

\noindent Fig. 4: Spectra of A (light curve) and B (heavy 
curve). Straight sections of the spectra are the locations of poorly
subtracted telluric lines. Absorption and emission lines (Table 3) are
labeled. 

\noindent Fig. 5: Background-subtracted spectra extracted from the
region of exposure 1 from 0\farcs64 to 1\farcs49 N of B (top to
bottom). From bottom to top, the ordinates of four successive spectra
are shifted down by 0.5, 1.0, 1.5 and 2.0 counts 
normalized by the continuum, relative to the
northernmost spectrum.

\noindent Fig. 6: Same as Figure 5, except for exposure 2.

\noindent Fig. 7: Same as Figure 5, except for exposure 3.

\noindent Fig. 8: Same as Figure 5, except for exposure 4.

\noindent Fig. 9: Calibration of the ``output'' FWHM of the fxcor
gaussian fit to the cross-correlation peak as a function of ``input''
dispersion $\sigma_v$ in \kms.  The solid (dotted) curve is the
calibration for the smoothed average Lick (average LRIS) template.
For comparison, the dashed curve shows the calibration for the average
Lick template without smoothing.  As expected, the use of the dotted
curve taken with LRIS only data would lead to smaller estimates of the
dispersion of G1.  The straight lines indicate the conversions of
FWHM to $\sigma_v$ for G1, for the average of our 20 useful exposures
(see the text).  For an ``input'' of $\sigma_v = 0$ \kms, each FWHM is
approximately equal to the FWHM of the cross-correlation peak of each
template with itself, given the spectral resolution in each case.

\noindent Fig. 10: Spectra of G1 (solid curve) and of the average smoothed
Lick template star (dashed curve).  Each spectrum was normalized by
its continuum for this comparison.  The template was convolved with a
gaussian with $\sigma = 279$ \kms, and G1 was blueshifted from $z =
0.356$ to rest.

\noindent Fig. 11: Spectra of the average Lick template star.  Both
spectra were normalized by their continua for this comparison.  The
solid curve is the original smoothed average template; the dashed
curve shows the result of a convolution of the original smoothed curve with a
gaussian with $\sigma = 279$ \kms.

\noindent Fig. 12: Variation of the velocity dispersion as a function
of the distance to the center of brightness of G1. Each point
represents one row extracted from our spectra, as explained in the
text.  Rows North (South) of the center of brightness of G1 are
represented by squares (triangles). The mean of the values is
indicated by the dashed line. Error bars indicate the standard error
in the mean for each row, which we estimated from our 20 rows
with useful exposures. 

\newpage

\section{TABLE CAPTIONS}

\noindent Table 1: Observations log with LRIS/Keck I, 1995 March 28.

\noindent Table 2: Absorption and emission lines in the spectra of 
\qp\ A, B.

\noindent Table 3: Cross-correlation results.

\noindent Table 4: Error budget for $\sigma_v$ estimates.

\newpage

\begin{table}[h]
\begin{center}
\begin{tabular}{clcrrccc}
\multicolumn{8}{c}{Table 1: Observations log} \\
\hline
\hline
Exposure & 
Object  &
Central $\lambda$& 
\multicolumn{1}{c}{Exposure} & 
\multicolumn{1}{c}{UT}&  
\multicolumn{1}{c}{airmass} & 
\multicolumn{1}{c}{PA} & 
\multicolumn{1}{c}{seeing FWHM}\\
         &         & 
\multicolumn{1}{c}{\AA} &  
\multicolumn{1}{c}{sec}    &   &       &
\multicolumn{1}{c}{$^{\circ}$ E of N} &
\multicolumn{1}{c}{$''$}\\
\hline
1   &    G1    &   7000    & 1500\ \ &  8:02:56 &  1.24 &  0\ \ & 0.90\ \ \\
2   &    G1    &   7000    & 1500\ \ &  8:33:53 &  1.25 &  0\ \ & 0.80\ \ \\
3   &    G1    &   7000    & 1500\ \ & 10:46:56 &  1.49 &180\ \ & 0.75\ \ \\
4   &    G1    &   7000    & 1500\ \ & 11:17:36 &  1.61 &169\ \ & 0.74\ \ \\
5   &    HD136711& 5000    &    4\ \ & 12:11:53 &  1.04 &  0\ \ & 0.54\ \ \\
6   &    HD132737& 5000    &    1\ \ & 12:00:00 &  1.03 &  0\ \ & 0.49\ \ \\
7   &    M81   &   5000    &   30\ \ &  5:36:59 &  1.67 &  0\ \ & 0.70$^1$\\
8   &    AGK2$+$14\,783& 5000&    2\ \ &  5:55:15 &  1.02 &  0\ \ &0.72\ \ \\
\hline
\end{tabular}
\end{center}
\bigskip
$\,^1$ We assumed the seeing FWHM for M81 was comparable to that for
AGK2$+$14\,783, because there were no unresolved objects in the
spectra that we could use to obtain an estimate.
%\label{log}
\end{table}
\begin{table}[h]
\begin{center}
\begin{tabular}{lrrrrr}
\multicolumn{6}{c}{Table 2:  QSO absorption and emission lines} \\
\hline\hline
 & \multicolumn{3}{c}{A} & \multicolumn{2}{c}{B}\\
Ion &$\lambda_{rest}$ &$\lambda_{obs}$& $W_\lambda$ &
$\lambda_{obs}$& $W_\lambda$ \\
 &   \AA\  &     \AA\  &    \AA\  &    \AA\  &     \AA\ \\
\hline
Fe II& 2599.4&  6216.4& 3.3& 6215.9& \ \ \ 3.5\\
Mg II& 2797.5&  6684.7& 5.3& 6684.0& \ \ \ 5.3\\
Mg II& 2802.5&  6701.8& 4.6& 6701.3& \ \ \ 4.6\\
Mg II& 2852.1&  6819.8& 0.6& 6819.8& \ \ \ 0.8\\
Mg II$^1$& 2797.5&  6757.3& $-$69.$^2$& 6758.2& $-$65.\\
\hline
\end{tabular}
\end{center}
%\label{abs}
$\,^1$ The only strong QSO emission line is that of Mg II 2797.5 \AA. 
$\,^2$ Negative equivalent widths correspond to emission lines.
\end{table}

\bigskip
\begin{table}[p]
\begin{center}
\begin{tabular}{clccr}
\multicolumn{5}{c}{Table 3:  Cross-correlation results} \\
\hline
\hline
Spectrum & Template & FWHM                     & $\sigma_v$ & $R-$value \\
         &          & \multicolumn{1}{c}{\kms} &\multicolumn{1}{c}{\kms}&\\
\hline
Exp. 1   & average &    696. &   285. & 13.\\
Exp. 2   & average &    723. &   300. & 12.\\
Exp. 3   & average &    661. &   267. & 15.\\
Exp. 4   & average &    660. &   266. & 11.\\
sum $1-4$  & HD126778 & 672. &   272. & 12.\\
sum $1-4$  & HR4435 &   679. &   277. & 12.\\
sum $1-4$  & HD172401 & 676. &   275. & 12.\\
row 1$^1$  & average &  680. &   277. & 11.\\
row 2  & average &      785. &   332. & 11.\\
row 3  & average &      723. &   300. & 12.\\
row 4  & average &      649. &   260. & 15.\\
row 5  & average &      653. &   262. & 14.\\
\hline
\end{tabular}
\end{center}
\bigskip
%\label{results}
\noindent $^1$ Rows are separated by $\sim 0\farcs213$, starting $\sim
0\farcs640$ North and ending $\sim 1\farcs49$, also North of QSO image B
(see Fig. 1).
\end{table}

\bigskip
\begin{table}[h]
\begin{center}
\begin{tabular}{lr}
\multicolumn{2}{c}{Table 4:  Error budget for $\sigma_v$ estimates.} \\
\hline\hline
error source &standard error\\
 &     \kms\ \ \ \ \\
\hline
choice of template star& 9\ \ \ \ \ \ \ \ \ \\
cross-correlation fitting width & 5\ \ \ \ \ \ \ \ \ \\
choice of absorption lines & 7\ \ \ \ \ \ \ \ \ \\
\hline
\end{tabular}
\end{center}
%\label{abs}
\end{table}

\newpage 

\clearpage
\begin{figure}
\plotone{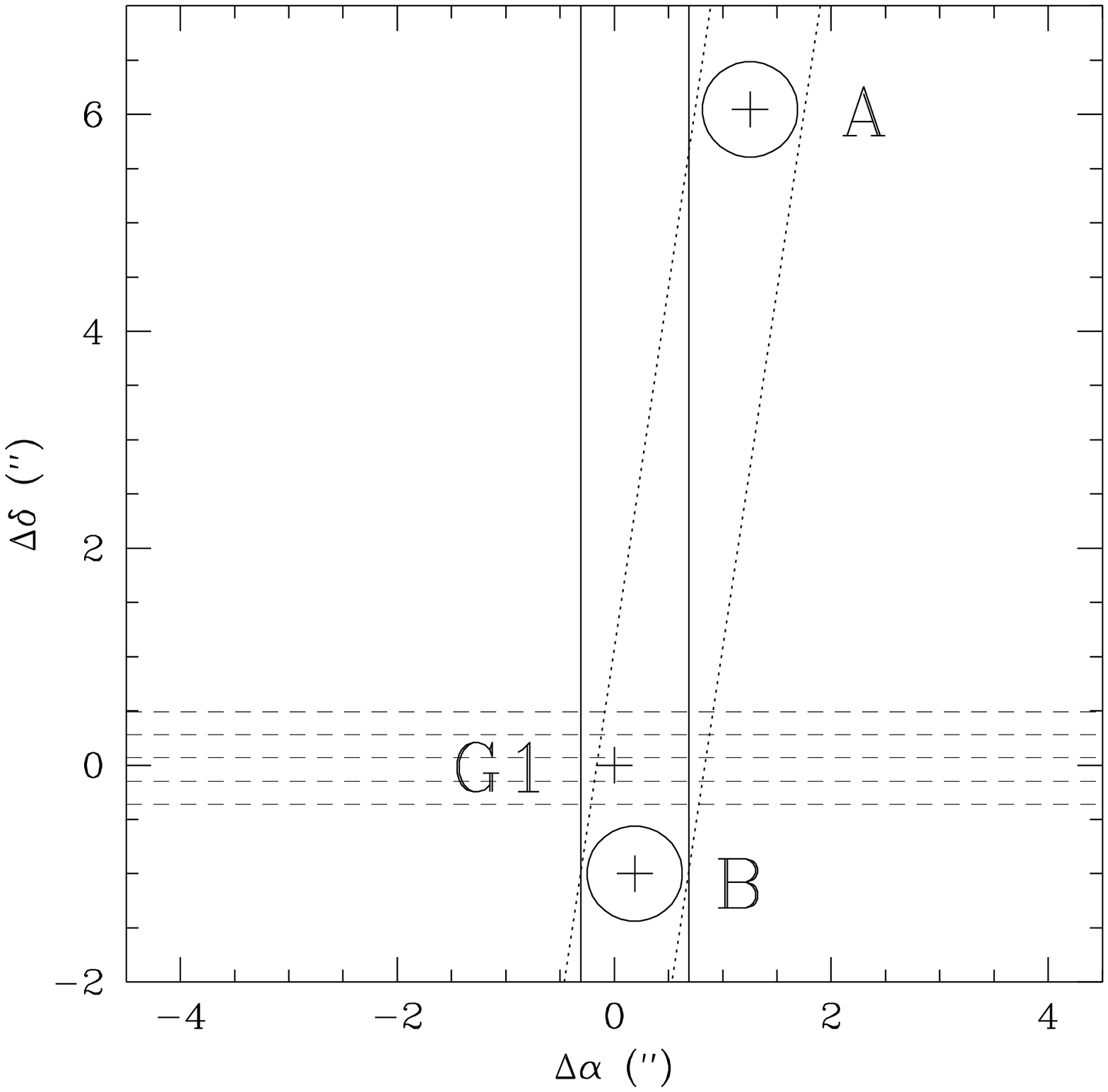}
\end{figure}

\clearpage

\begin{figure}
\plotone{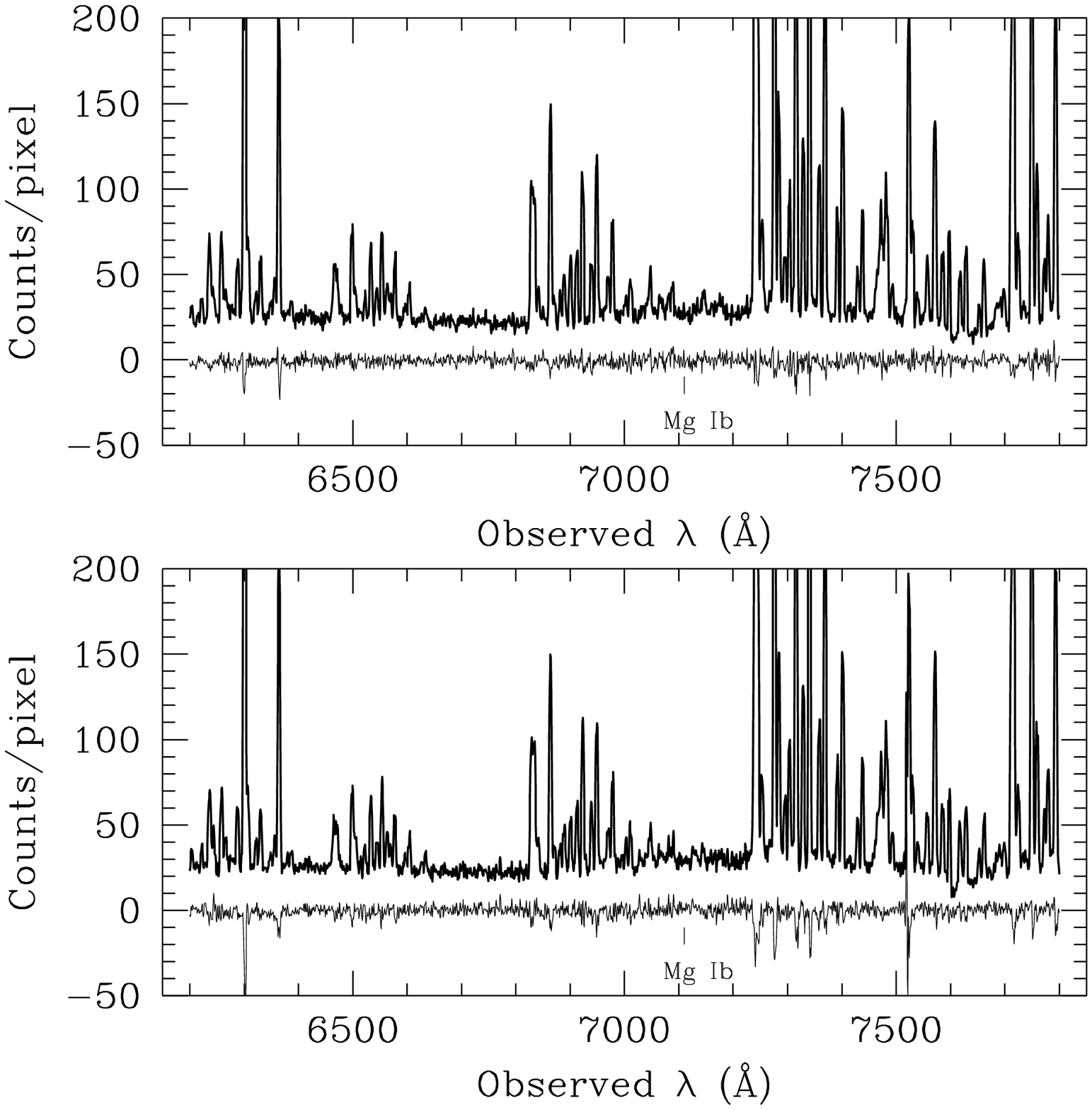}
\end{figure}

\clearpage

\begin{figure}
\plotone{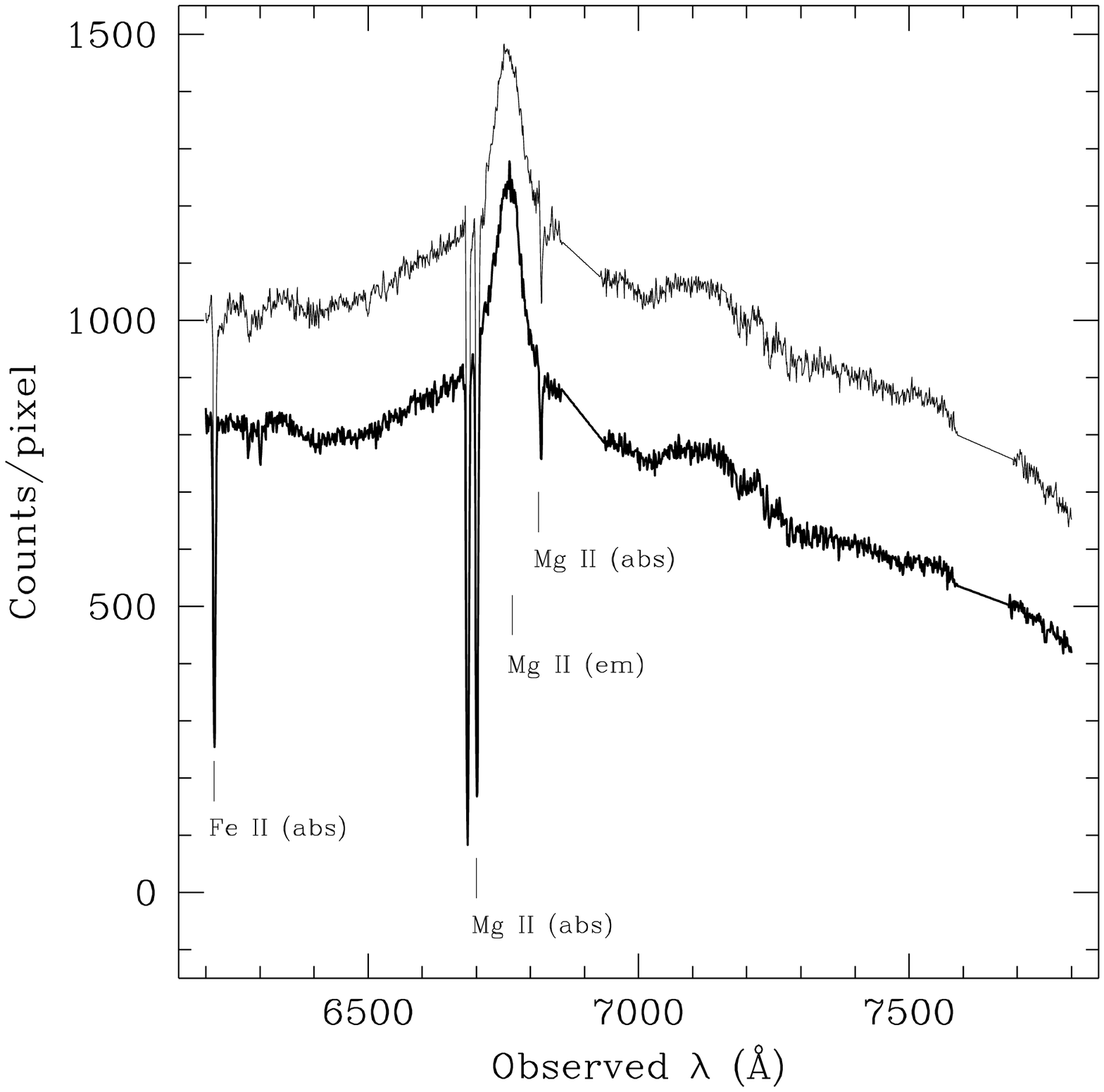}
\end{figure}

\clearpage

\begin{figure}
\plotone{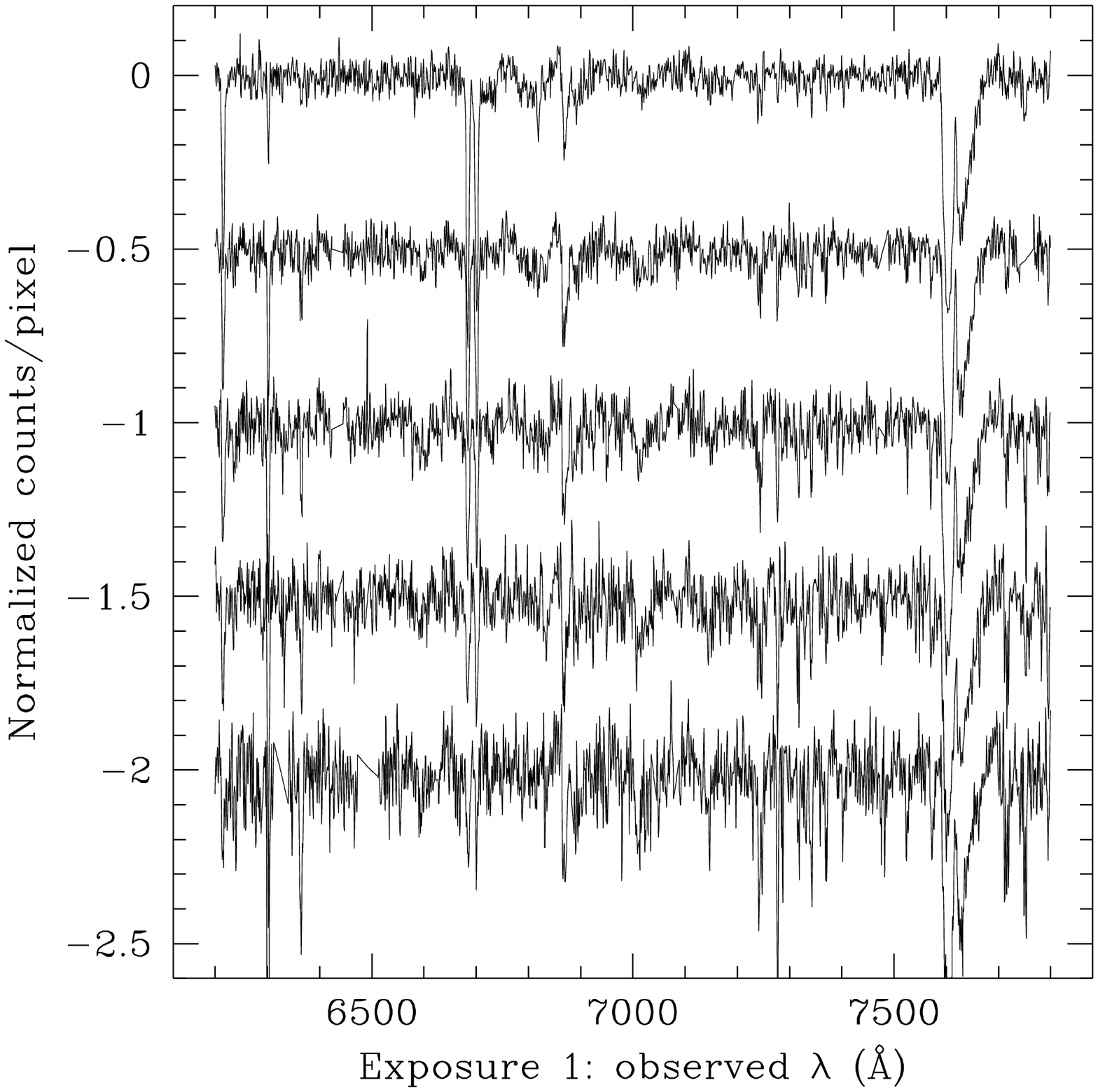}
\end{figure}

\clearpage

\begin{figure}
\plotone{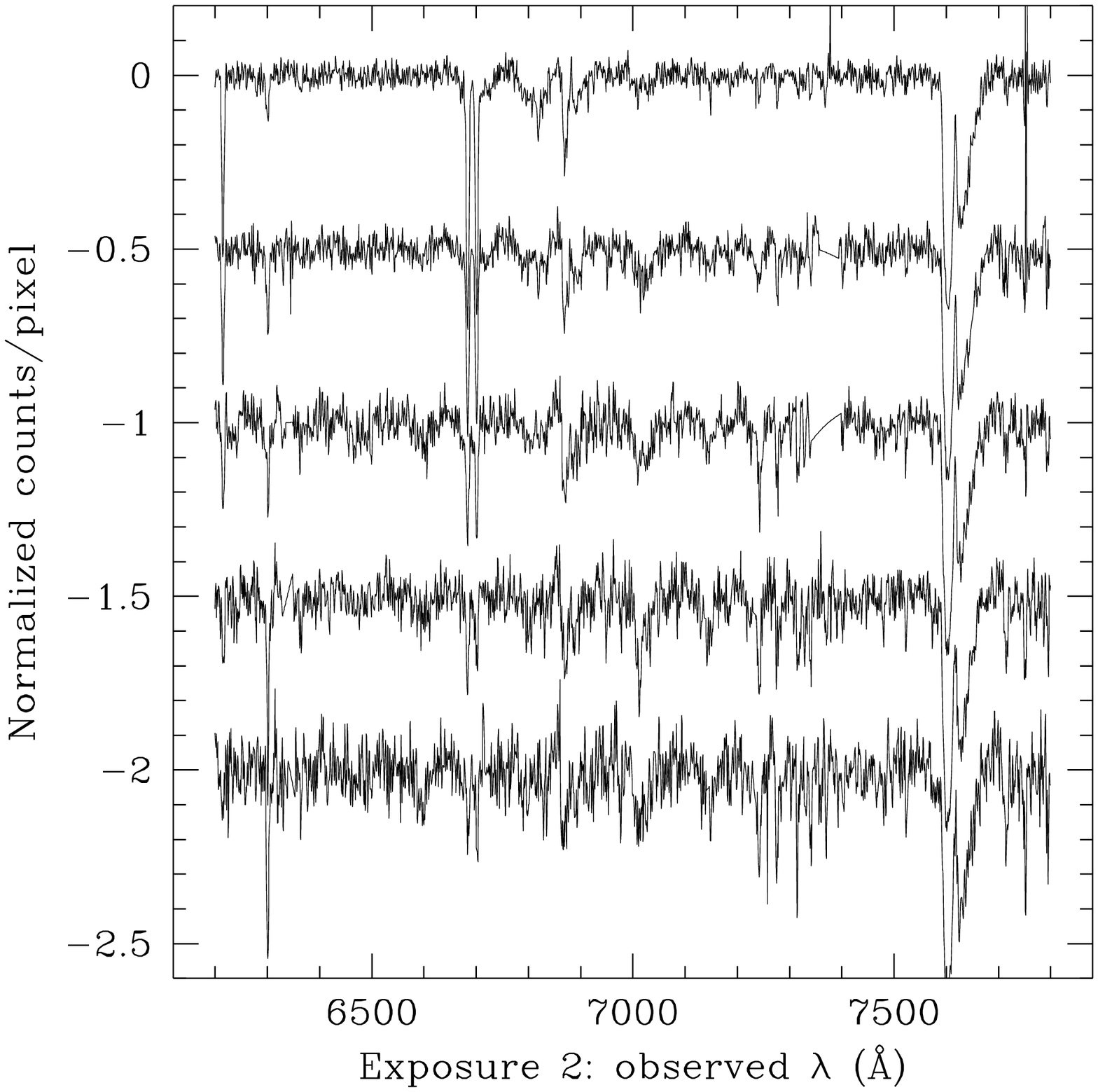}
\end{figure}

\clearpage

\begin{figure}
\epsscale{1.0}
\plotone{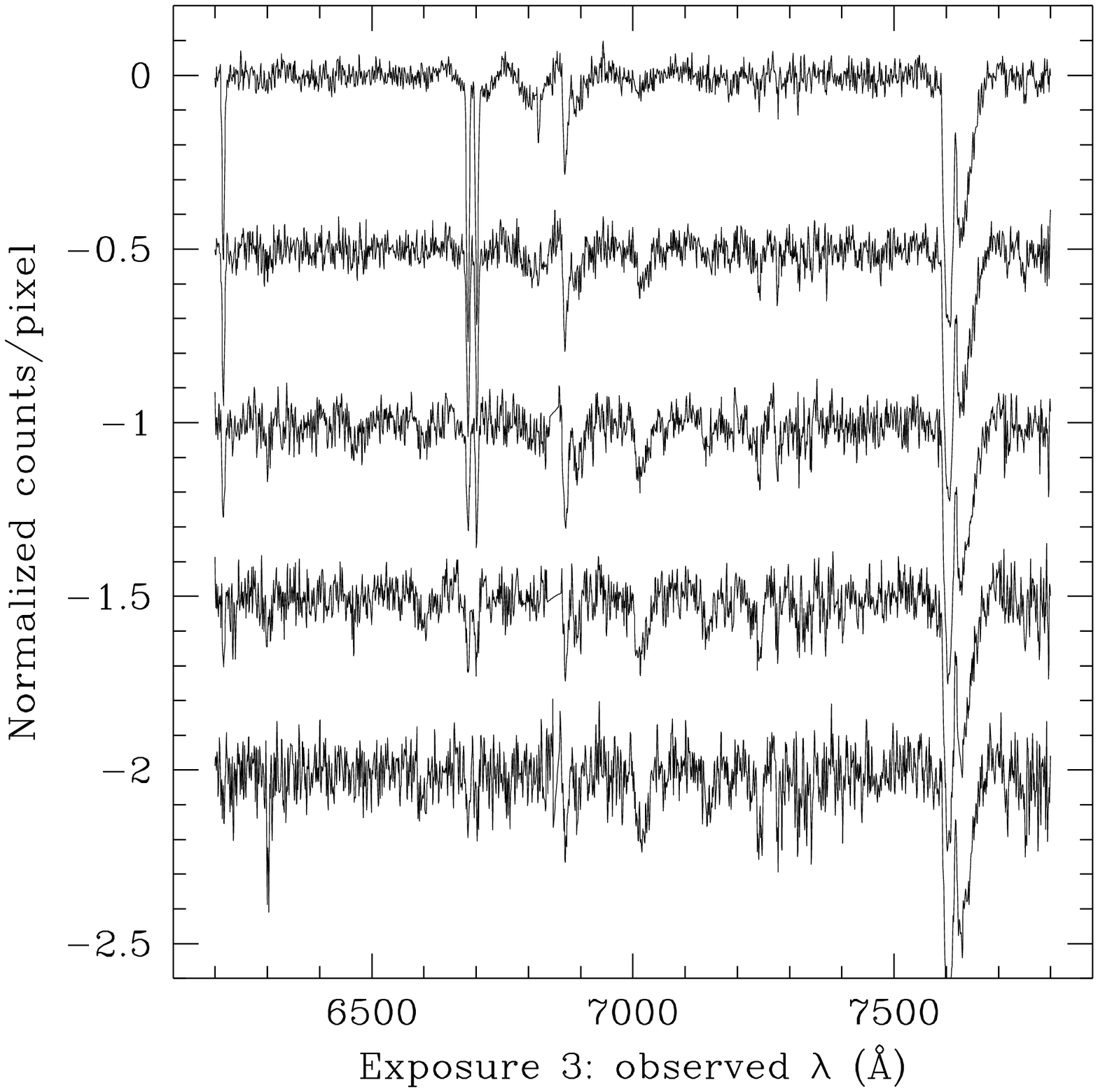}
\end{figure}

\begin{figure}
\epsscale{1.0}
\plotone{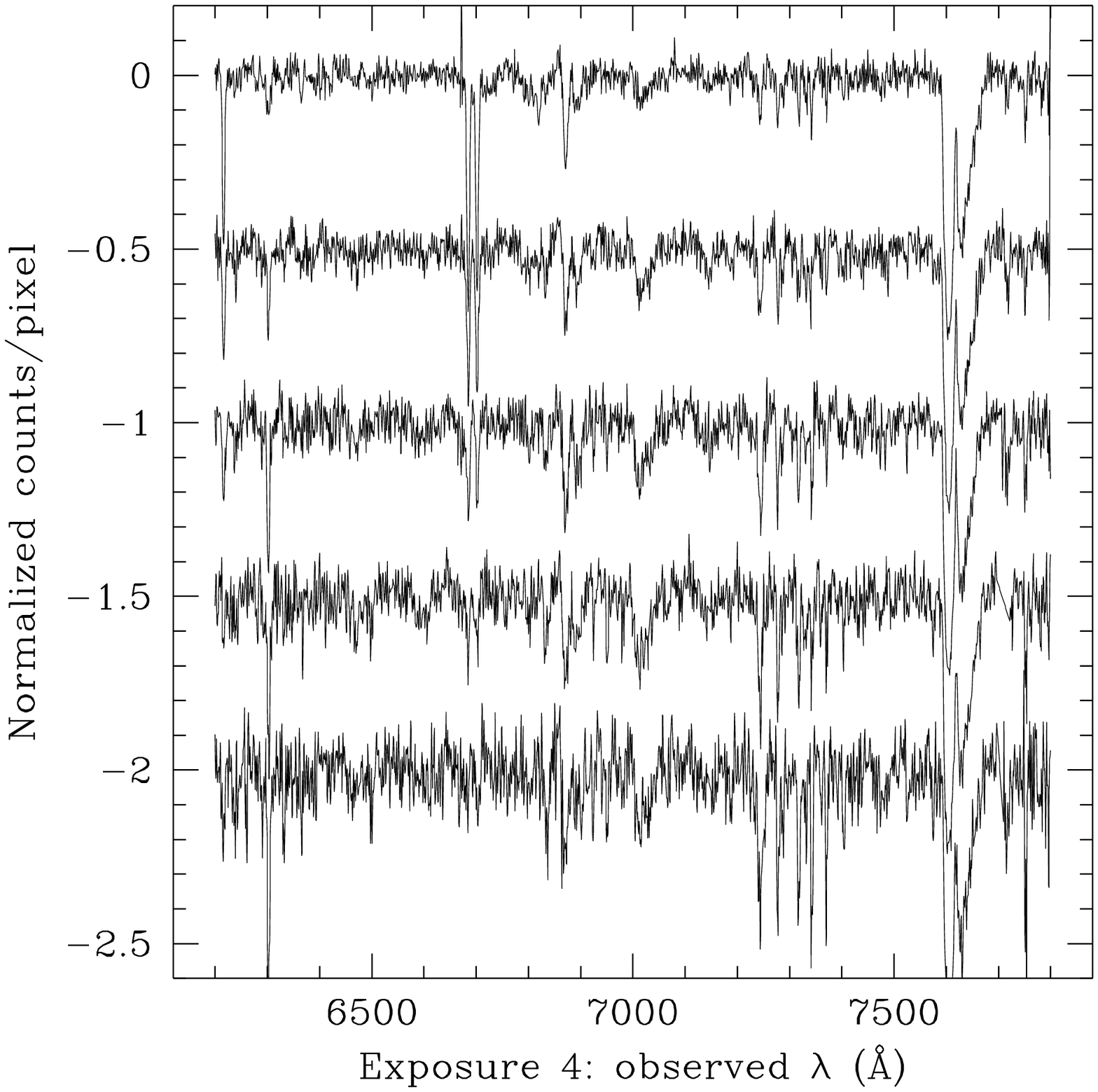}
\end{figure}

\begin{figure}
\epsscale{1.0}
\plotone{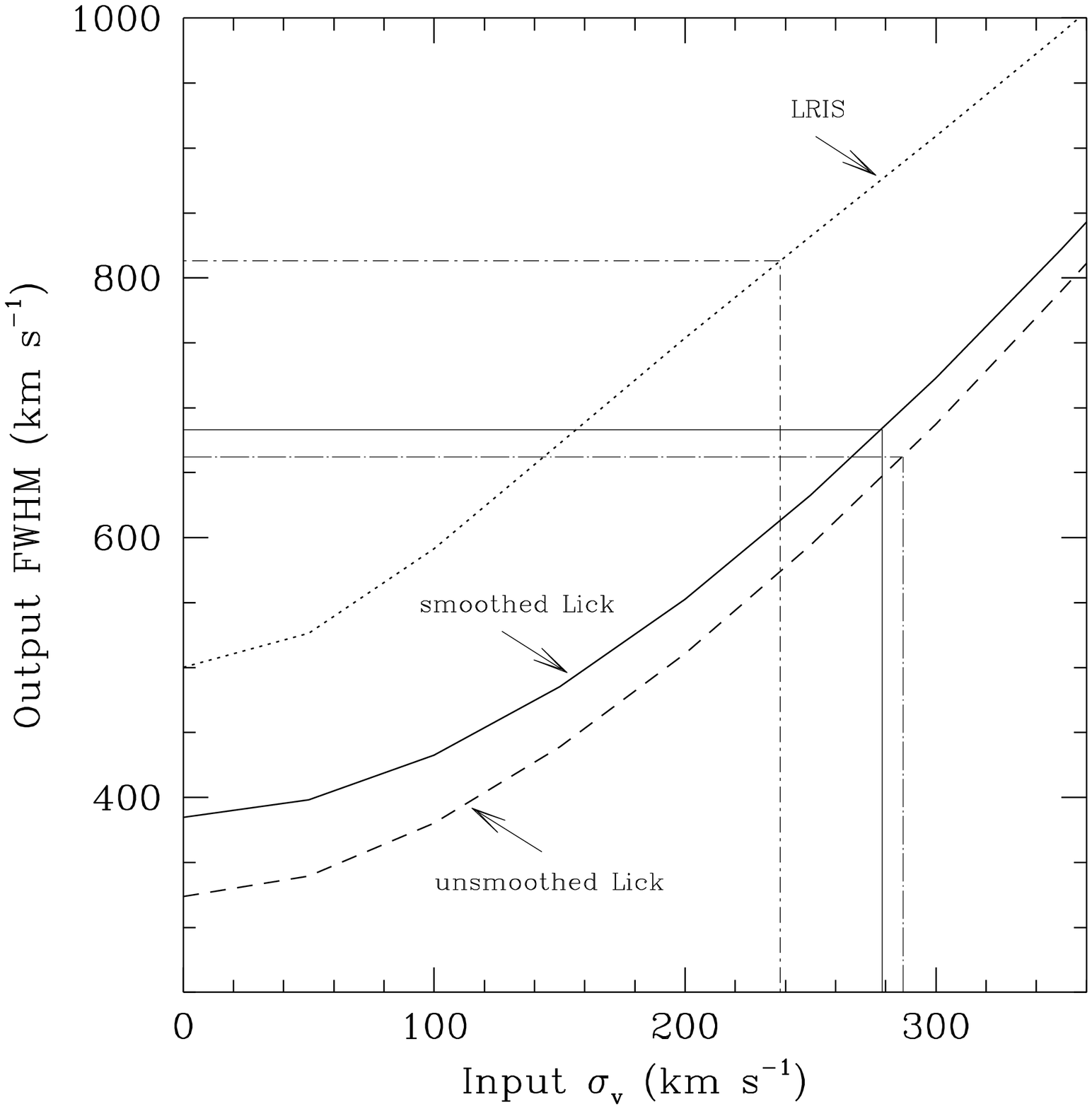}
\end{figure}

\begin{figure}
\epsscale{1.0}
\plotone{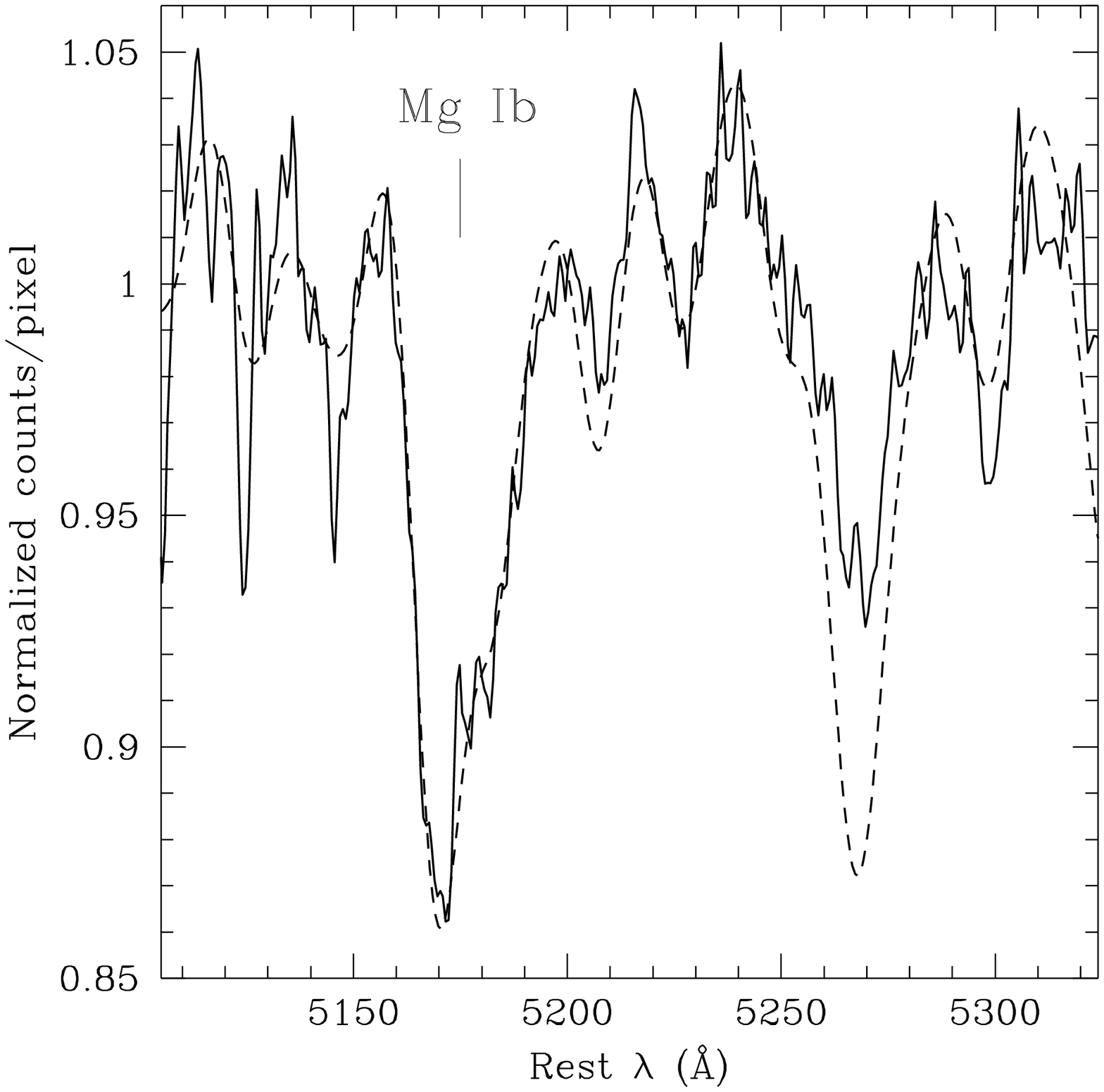}
\end{figure}

\begin{figure}
\epsscale{1.0}
\plotone{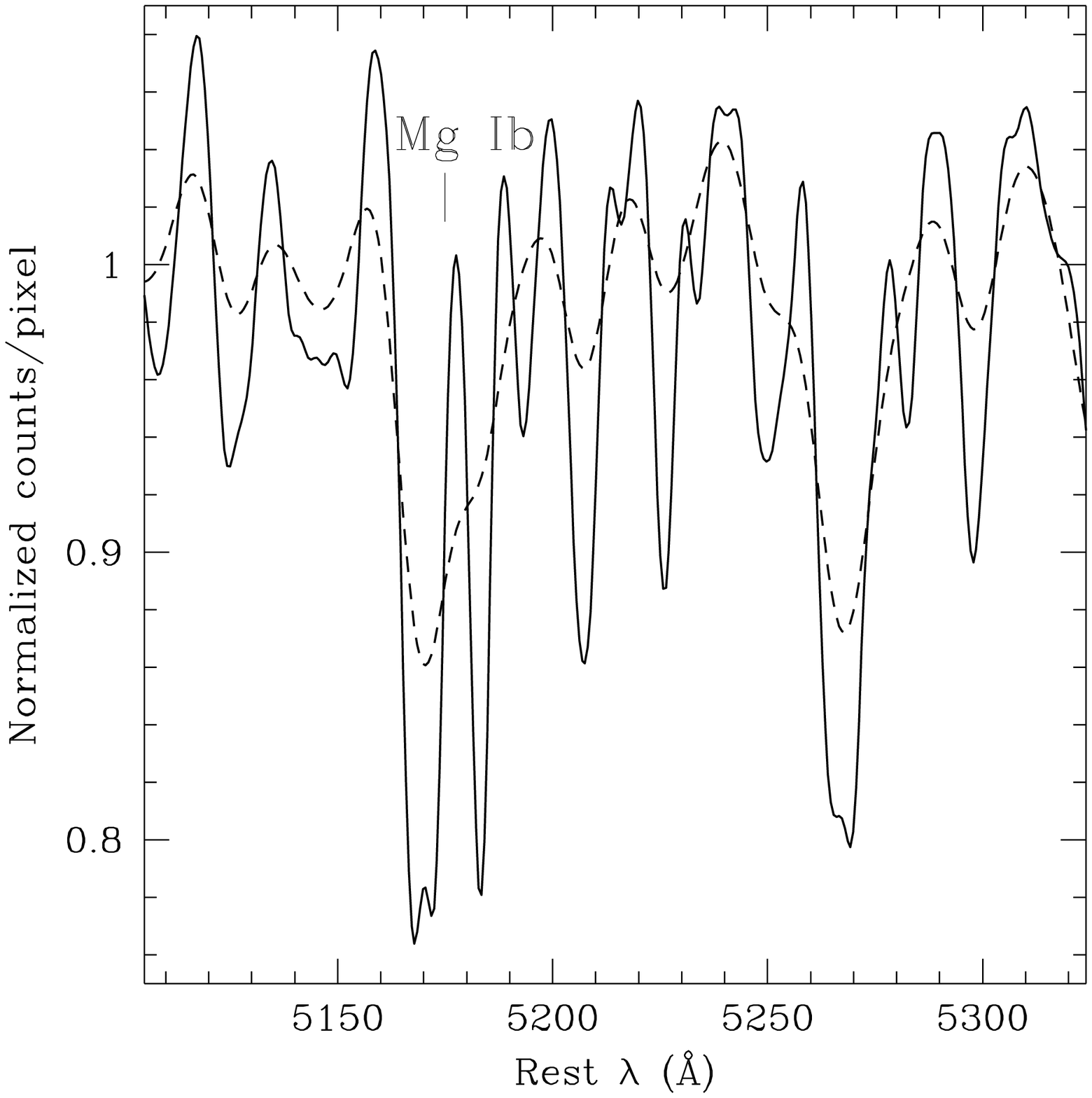}
\end{figure}

\begin{figure}
\epsscale{1.0}
\plotone{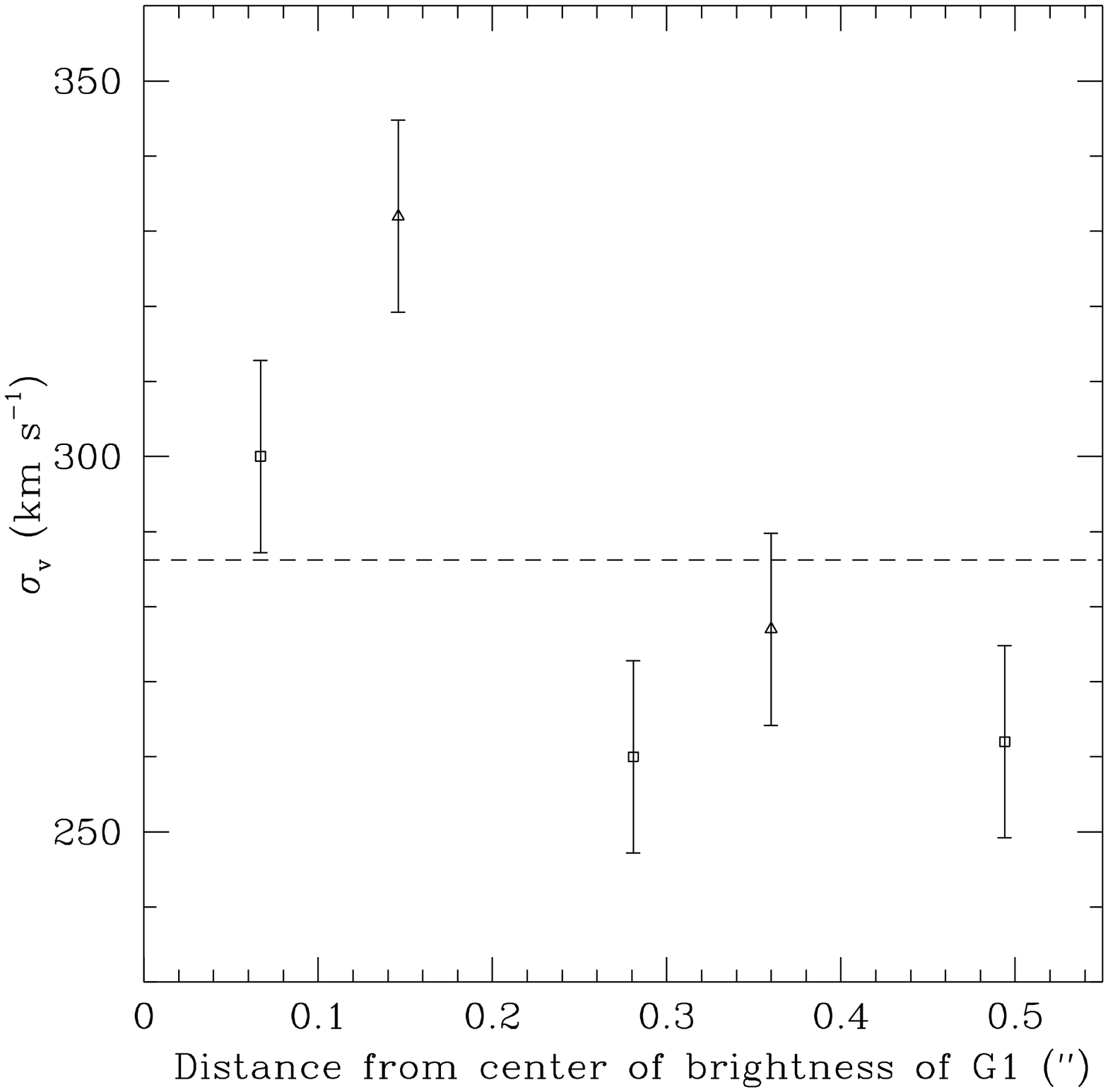}
\end{figure}

\end{document}